# Looking Through Glass: Knowledge Discovery from Materials Science Literature using Natural Language Processing

*Vineeth Venugopal[1,*], Sourav Sahoo[2], Mohd Zaki[1], Manish Agarwal[3], Nitya Nand Gosvami[2], N. M. Anoop Krishnan[1,2,*]*

[1]Department of Civil Engineering, Indian Institute of Technology Delhi, Hauz Khas, New Delhi 110016, India
[2]Department of Materials Science and Engineering, Indian Institute of Technology Delhi, Hauz Khas, New Delhi 110016, India
[3]Computer Services Center, Indian Institute of Technology Delhi, Hauz Khas, New Delhi 110016, India

[*]*Corresponding authors: V. Venugopal (vinven7@gmail.com), N. M. A. Krishnan (krishnan@iitd.ac.in)*

**Most of the knowledge in materials science literature is in the form of unstructured data such as text and images. Here, we present a framework employing natural language processing, which automates text and image comprehension and precision knowledge extraction from inorganic glasses' literature. The abstracts are automatically categorized using latent Dirichlet allocation (LDA), providing a way to classify and search semantically linked publications. Similarly, a comprehensive summary of images and plots are presented using the 'Caption Cluster Plot' (CCP), which provides direct access to the images buried in the papers. Finally, we combine the LDA and CCP with the chemical elements occurring in the manuscript to present an 'Elemental map', a topical and image-wise distribution of chemical elements in the literature. Overall, the framework presented here can be a generic and powerful tool to extract and disseminate material-specific information on composition–structure–processing–property dataspaces, allowing insights into fundamental problems relevant to the materials science community and accelerated materials discovery.**

**Introduction**
Overwhelmingly large amount of knowledge generated through scientific enquiry is stored as unstructured data in the form of texts and images. These range from expository archives such as books, journals, and dissertations to condensed representations such as handbooks and manuals. Materials science, being a highly interdisciplinary area, commands a large repository of scientific publications. However, only a limited fraction of this knowledge is collected and curated in the form of structured data, for example, a database of composition–structure–property relationships. The information on material science is increasingly siloed and simply too large for the efficient utilization of any one individual or group. Just as in all other branches of science, the materials community is afflicted by the curse of knowledge incommensurate with the available information[1,2]. Thus, the accessibility to the vast majority of knowledge in the literature is limited, as it: (i) is time consuming to manually read and analyze the texts and images, and (ii) requires a domain expert to understand, interpret, and summarize the information.

Recent advancements in natural language processing (NLP) provide a promising solution to this problem through the automation of text comprehension, querying, and knowledge extraction from scientific texts. NLP has been applied extensively to scientific literature specifically in biological sciences for more than two decades[3-5]. A biomedical-specific language model, namely, BioBERT[4], which is used extensively for biomedical text mining stands as a testimony to the advances and contributions of NLP in biological sciences. In contrast, the applications of NLP to materials science remain sparse[6,7]. Similar to biological sciences, the study of materials present some unique challenges to the direct application of NLP to mine text data due to the domain-specific jargons and lack of uniform conventions in scientific writing[7,8]. Despite these challenges, recent studies have shown that NLP can indeed be used to address some open challenges in materials science such as novel materials discovery[6], unraveling synthesis pathways[9], and extracting composition–property databases[10].

Cole et al[11-13] have demonstrated the automated generation of databases for magnetic[14] and battery materials[10] using ChemDataExtractor[15], which has also been used in predicting phase diagrams[16]. Olivetti et al[7,9,17] have used NLP together with artificial neural networks to predict synthesis parameters of inorganic oxides[9,18,19] and in extracting the properties of zeolites[20] and cementitious materials[21]. Jain et al[6] have demonstrated the use of word vectors in converting semantic queries to vector algebra and extended the method to the prediction of thermoelectrics. Ceder et al[9] have shown the extraction of automated synthesis recipes of inorganic oxides through a semi-supervised approach. Recently, Matscholar[8] has been introduced as a comprehensive material science search and discovery engine that is able to automatically identify materials, properties, characterization methods, phase descriptors, synthesis methods and applications from a given text through a custom built named entity recognition (NER) system. These developments suggest that artificial intelligence approaches using NLP can be a promising route to condense and represent knowledge in materials science leading to novel materials discovery and development.

Very few studies have, however, focused on extracting information related to images and plots in literature. The adage, "a picture is worth a thousand words", is even more relevant to scientific literature as images hold the most crucial information related to scientific hypothesis and theories. Till date, there has been no framework that allows direct search or compilation of images presented in scientific literature. Further, the images of a manuscript should be read in conjunction with the text to understand the context. While many of the applications of NLP in material science have focused on extraction and processing of textual information, no effort has been made thus far to connect these textual information with the images and plots to allow knowledge dissemination in a holistic manner.

Here, we demonstrate a comprehensive NLP framework that extracts information from a large corpus of text and images to provide highly specific, nuanced, and automated exploration of materials science literature. Specifically, we analyze the texts and images from around 100,000 research articles in the area of glasses, an archetypical disordered material. Glasses are one of the most common and widely used among engineering materials with uses spanning architectural, functional, and biomedical applications[22,23]. Recently, machine learning (ML) approaches have been used to develop predictive models for optical, electronic, and mechanical properties of glasses[22,24-34]. Several recent works have shared composition–property databases along with the trained ML models[29,30,33,35]. For instance, the software package, Python for Glass Genomics (PyGGi), has a large composition–property database, ML models for predicting nine key properties and an optimization framework for targeted glass discovery[36]. These models, however, have relied on existing databases for their training and analysis[37], and hence have been restricted to parameter predictions through regression models. It is well known that the properties of glasses, a non-equilibrium state, are not just a function of composition, but are also fundamentally influenced by the processing history and testing conditions[23,38-40].

Through a combination of NLP algorithms, chemical entity extraction protocols, and visualization tools, we show that the answers to very specific questions on glass literature can be answered. These include material/property specific questions as well as broader community issues such as:
1. What are the common synthesis methods for a given glass? Eg: What is the most common synthesis method for optical glasses?
2. What chemical elements have been used in LEDs?
3. Where is Americium used in glasses?
4. What glass compositions have been studied with an AFM?
5. Are more of the papers published in glass science theoretical as opposed to experimental?
6. What glass compositions have been studied for a given property? Eg: What glasses are known to be fluorescent in the ultra violet?

Overall, the generic framework developed here allow highly specific exploration of scientific literature using the abstracts, text and image captions of publications.

**Methods**
The CrossRef metadata API was used to query existing literature databases using keywords specific to the glass community. These include (1) descriptive application identifiers such as 'chalcogenide glasses', 'bioactive glasses', 'laser glasses', 'optical glasses,' etc (2) property/processing terms such as 'glass transition temperature', 'oxide glasses', 'optical luminescence', etc, and (3) conjugated keywords such as 'glass mechanical properties', 'glass-dissolution', 'glass AND fracture' etc. This query returned an initial list of over 6 million DOIs out of which the full texts of 600,000 articles were downloaded using the Elsevier Science Direct API. A custom XML parser was written to extract specific sections of the article including the metadata, abstract, images, image captions, and individual sections identified by their headings.

To understand the distribution of topics in the downloaded database, a natural language processing (NLP) algorithm called the Latent Dirichlet Allocation (LDA) was used to identify the number of distinct 'topics' in the corpus where a topic is defined as the set of words with the highest probability of occurrence in a document belonging to the topic. The LDA plot is presented in the supplementary data. It is seen that while some topics are relevant to the scientific literature on glass, many are from the intersection of glasses with peripheral topics such as women's health, economy, and environment. In fact, one topic only includes non-English articles identified by the European vowels 'un', 'es', 'le' etc. The full texts of articles belonging to topics of little relevance to the materials science literature on glasses were removed, along with editorial notes, commentaries, book reviews, retractions, and conference proceedings. The use of LDA is thus demonstrated to greatly aid the curation of topic specific text databases – a non-trivial effort by any other means.

The remaining database was further refined by the use of a machine learning classifier model that performed a binary classification of the article abstracts into 'relevant' and 'non-relevant'. For this, 3000 randomly selected abstracts were manually tagged as being 'glass' and 'not-glass'. Using the Python SciKit library, a logistic regression classifier model was found to have an accuracy of 86 % and a recall of 76 % on the classification task. The regression model was found to outperform other text classification algorithms such as Naïve Bayes and Random Forests. The details are the models are given in the supplementary materials. Finally, the binary classifier categorized 94,207 articles as being relevant to the material science study of glasses. All further natural language processing driven analyses were done only on this text corpus. The DOIs, Titles, Journals and Authors of these articles are given in additional data.

**Caption Cluster Plot**

The final corpus contains a total of 106,238 figures and their captions. With the help of the Stanford NLTK package the caption texts were tokenized after removal of punctuations, numerals, and stop words. These tokens, which include words of the English language, chemical symbols and abbreviations, form the corpus dictionary of size (N) which determine the number of dimensions of the vector space for the Caption cluster plot. Each caption is mapped to a unique vector in this vector space by calculating the Term Frequency - Inverse Document Frequency (TFIDF) of every word in the caption. TFIDF is a statistical count that reflects the relevance of a word in a document and is a common vectorization technique for text mining and information retrieval. For a term "t" appearing in "d" documents within a collection D:

$$\text{TFIDF} = tfidf\,(t, d, D) = tf\,(t, d) \times idf\,(t, D)$$

where

$$tf(t, d) = \begin{cases} 1 \text{ if } t \text{ is present in } D \\ 0, otherwise \end{cases}$$

$$idf\,(t, D) = log \frac{N}{d[D:t[d]}$$

The cosine distance of all the captions from each other is calculated and used as the metric for T-Stochastic Nearest Neighbor Embedding (TSNE), which projects the vectors into a two dimensional plane so that vectors with the highest cosine similarities group together. Finally, each caption is assigned a unique label corresponding to the type of image it represents, such as SEM, XRD, TEM, Luminescence, Fracture, etc. through rule based string search. The color of the pixel in the TSNE plot is determined by its label. The TFIDF vectorization and the cosine metric ensure that the geometric distance between pixels on the 2D plot correlate with the semantic similarity of captions, and therefore that identical images cluster together.

**LDA Plot**
The 94207 abstracts are further categorized using LDA. The optimum number of topics was found by the coherence plot which was found to converge after 15 topics. After 500 passes, the LDA algorithm identified the topics listed in Fig 1(c). Similar to the caption cluster plot, each abstract was tokenized, vectorized, and plotted in 2D using TSNE. The coloring of the pixels is based on the topic number identified by the LDA. Once again, the pixels are seen to cluster strongly based on color, indicating that abstracts with similar lexical content have been grouped together by the algorithm.

**Extraction of Chemical species**
ChemDataExtractor was used to identify and extract the individual chemical species from every abstract. This includes individual chemical elements and compounds identified by their symbols, names and chemical formulae. A custom python script is used to extract the individual chemical elements from the most frequent of these compounds.

**Elemental maps**
If an element X is identified as being present in an abstract, the pixel corresponding to the abstract in the LDA plot is given a different color. This allows the mapping of elemental compositions as identified in the step above to the information content present in the LDA plot. Similarly, the elemental information and captions are correlated by merging the caption cluster plot and the LDA. A caption pixel is given a different color if the abstract of the text that the caption belongs to contains the element.

**Results**
To demonstrate the proposed approach, we downloaded more than 600,000 research articles, full texts and images, related to the keyword "oxide glasses" and "materials science" using the CrossRef metadata query API[41] and the Elsevier Science Direct API[42]. Following this, supervised learning was performed on the abstracts of the manuscripts to filter the relevant manuscripts. (see Methods for details). Abstracts are the most information dense organ of a scientific paper, containing information on the material under study, property being explored, and characterization/synthesis methods being employed in service of the investigation. As such, they are unlikely to contain spurious information or refer to materials or properties not mentioned in the text. This specificity makes an abstract the most useful part of a text, and it is therefore not surprising that many NLP studies on materials have only taken paper abstracts as the input[6]. Based on the supervised learning, we obtained approximately 100,000 research articles relevant to the topic. While not an exhaustive list, the total number of articles downloaded are in the same range as the number of texts identified in other comprehensive literature surveys on glass[43,44].

An unsupervised NLP algorithm called latent Dirichlet allocation[45] (LDA) was used to automatically classify the corpus into fifteen 'topics', where each topic is defined by the set of words that have the highest probability of occurrence within the topic. LDA allows a rapid and efficient organization of the text corpus with minimal human supervision—a capability provided by no other automated tool available today. The categories generated by LDA is visualized in the LDA plot in Fig 1(a). Each abstract in the corpus is vectorized using Term Frequency–Inverse Document Frequency[46] (TFIDF) which maps each document in the corpus to a unique vector in a higher dimensional space. t-distributed stochastic neighbor embedding (t-SNE) projects these vectors to a 2D plane such that vectors with the highest cosine similarity group together. The color of a pixel is determined by the topic number assigned to it by LDA.

It is seen immediately from Fig. 1(a) that the pixels group together by color, which demonstrates the agreement between two different organizational paradigms—vectorization followed by clustering, and LDA. The LDA plot is a graphical representation of the entire field of glass literature and succinctly summarizes the details mentioned earlier. Note that the descriptive label is assigned to these topics by a human expert that maps the automatically generated lexical probability distribution to established categories in glass literature. For example, the words with the highest probability of occurrence in Topic 11 are 'er', 'yb', 'emission', 'doped', 'luminescence', 'nd', and 'tm'. This suggests that the topic is related to the luminescence of glasses doped with rare earth ions and hence is labelled as 'Rare Earth glasses'. The distribution of abstracts into the identified topics is shown schematically in the histogram in Fig. 1(b). The descriptive labels for all the other topics are similarly identified and listed in Fig. 1(c).

A mere visual inspection confirms that Topic 5–thin films–is the single largest group followed by topics 13 and 6. The most common set of articles related to 'Thin films' includes both the luminescence properties of oxides such as ZnO on glass substrates as well as the studies of transparent glasses on a thin film geometry. This is followed by 'glass ceramics', 'methodological studies' of glasses (including both theoretical and modeling studies), and 'non-linear optical properties of glasses'. In general, the list is found to be comprehensive covering all facets of glass literature in terms of applications, properties, and ingredients including bioactive glasses, dielectric glasses, nanomaterials, chemical and electromagnetic properties. Other categories such as mechanical and failure studies of glasses are seen to be subsumed within these broad topics. Note that a more detailed classification can be further performed to obtain the subtopics by performing LDA recursively on each of the topics separately (see Supplementary materials).

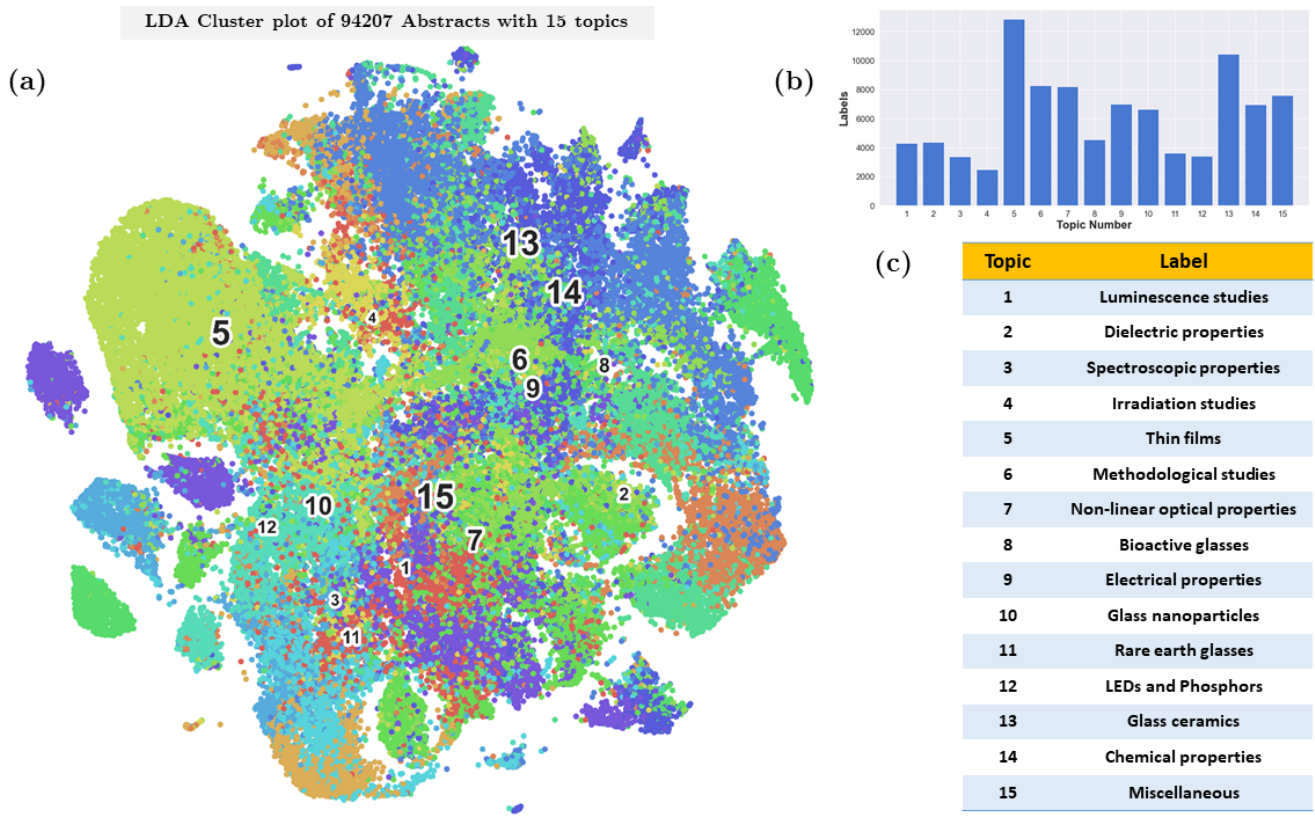

**Figure 1. LDA plot of the abstracts. (a)** The latent Dirichlet allocation (LDA) plot present the clusters of vectorized abstracts colored based on topics identified by LDA. (b) The number of abstracts on each topic as identified by LDA. (c) The descriptive label assigned to LDA topics by a human expert.

The corpus also contains a collection of 106,238 figures and their captions. It is well known that the information content of scientific articles is expressed mostly through graphics. These images and their corresponding captions therefore provide a technical summary of the documents that they belong to. The Caption Cluster Plot[47] (CCP) shown in Fig. 2(a) is a graphical representation of the information contained in all the captions, grouped by their semantic similarity using NLP. The captions are tokenized and vectorized using TFIDF and t-SNE, as explained earlier for the LDA plot. The pixels are colored based on categorical keywords identified by a human expert as explained in the methods section. Finally, these labels are positioned at the median (x,y) positions of the corresponding pixels, with the size of the label proportional to the number of images in that category.

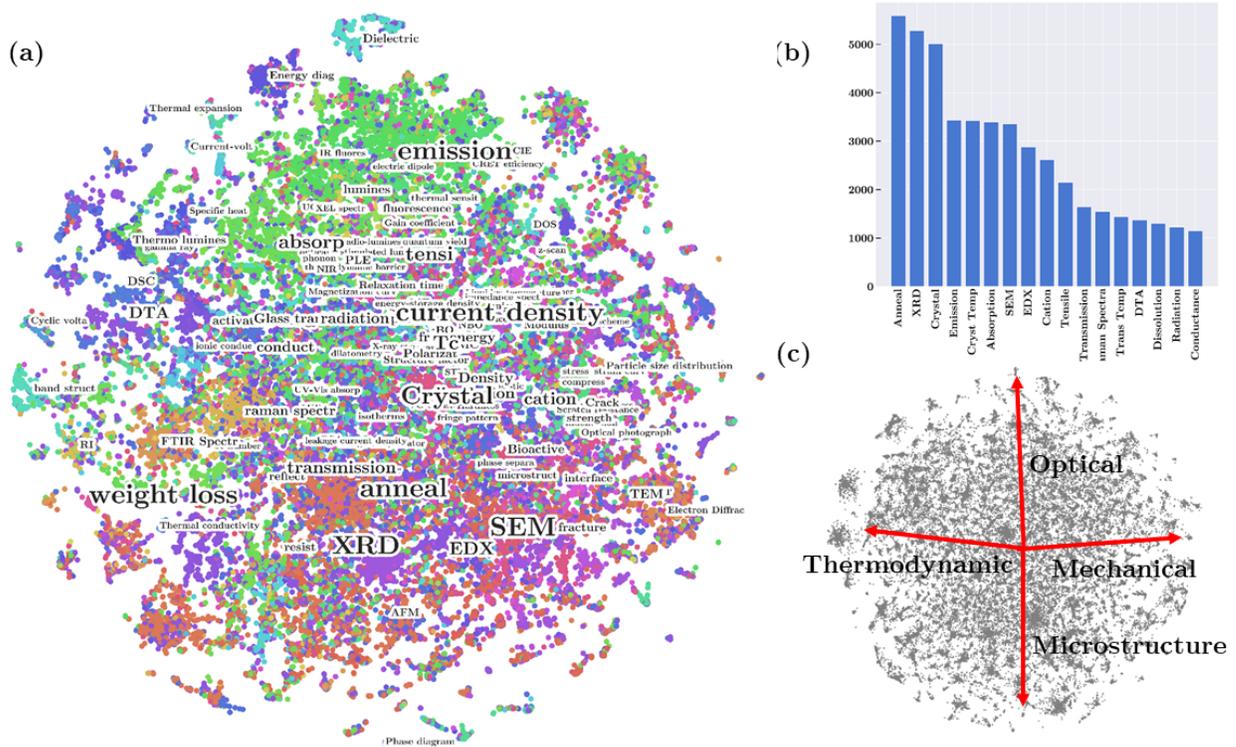

**Figure 2. Caption cluster plot. (a)** The caption cluster plot presents the clusters of vectorized captions colored by preselected keywords**. (b)** The labels with the highest number of counts in the caption database**. (c)** A grayscale image of the plot showing the four ontological axes.

Fig. 2(b) shows the distribution of captions with respect to the topics for each image type identified in the CCP. We observe that 'Anneal', Xray Diffraction (XRD), 'Crystal', 'Emission' and 'Crystallization temp' (Tc) are the most common types of images in glass literature, underlining the importance of thermal synthesis routines, microstructure and optical characterization methods in glass science. Captions representing similar types of images are found to cluster next to each other. For example, microstructure measurements such as Scanning Electron Microscopy (SEM), XRD, Atomic Force Microscopy (AFM), Energy Dispersive X-ray Spectroscopy (EDX), and Transmission Electron Microscopy (TEM) all cluster at the lower half of the caption cluster plot while optical properties such as emission, absorption, fluorescence, and luminescence are found at the very top. The use of TFIDF as the vectorization algorithm ensure that labels, and hence the captions representing images, are found to group together organically, based on their semantic similarity.

Interestingly, the CCP is found to have 4 distinct axes that capture complementary information on glass literature. These are the Optical, Mechanical, Microstructural and Thermodynamic axes as shown in Fig. 2(c). Optical axes is represented by images related to 'emission', 'luminescence', 'fluorescence' to name a few. Images that study mechanical properties such as 'crack', 'stress-strain', 'strength' and 'compressibility' are found closest to the Mechanical axes, while thermodynamic properties such as glass transition temperature, activation energy, specific heat, Differential Scanning Calorimetry (DSC), and Differential Thermal Analysis (DTA), to name a few, lie along the Thermodynamic axis. Finally, the images related to 'XRD', 'EDX', 'anneal' and 'Crystal' fall along the microstructure axis. This observation can be formalized by computing the Euclidean distance between caption labels and the four axes. The thermodynamic properties are closest to the Thermodynamic axes, while the optical properties are most proximate to Optical axes etc. This observation is a direct result of using the TSNE algorithm based on the cosine similarity of caption vectors, which ensures that the Euclidean distance between pixels is a measure of the semantic similarity of the underlying captions.

Certain captions do not belong to any one axes but are found to be equally spaced from two or more of them. For example, 'Fracture' and 'Interface' are located at nearly equal distances from the 'Mechanical axis' and 'Microstructure axis'. These terms represent microstructural features that are critical determiners of mechanical properties. The captions likely contain terms that relate equally to both axis categories, justifying their position in the CCP. The position of cluster labels is an indicator of their relative co-occurrence frequency, thereby showing that 'anneal' is strongly related to 'Crystal' (and variations thereof including 'crystalline' and 'crystal structure') while 'PLE' is more related to 'absorption' than to 'emission'. At the same time, the proximity of 'bioactive' to the mechanical axis suggests that many experiments on bioactive materials pertain to the measurement of strength and hardness thereby bringing the captions into semantic convergence. Overall, we observe that the CCP provides invaluable insights into the contents of these figures, which when combined with a predefined ontology solve many of the problems stated earlier. In particular, the analysis of captions through CCP provides a visual tool that rapidly summarizes the entire field of glass literature, allowing a user to quickly comprehend the trends, themes, and common characterization methods in the community.

Next, the python library, ChemDataExtractor, was used to automatically extracts chemical species—including names of compounds, chemical formulae and symbols—from the abstracts. The chemicals that occur with the highest frequency in the database are given in the Supplementary material. The chemical names and symbols were standardized following which the chemical elements present in each compound was separately identified. This creates a binary marker for each element such that if the element is mentioned in the abstract, the marker assumes the value 1 and 0 otherwise. The LDA plot is redrawn such that only if the abstract contains the marker for an element is the corresponding pixel colored. Similarly, if a caption is drawn from a text wherein the element is contained in the abstract, the pixel is marked with a color in the CCP. The results are the Elemental maps given in Fig. 3, where the images at the top are elements mapped to the caption cluster plot and the images at the bottom are maps of LDA. The elemental maps provide a direct visual representation of the distribution of elements in glass literature. The juxtaposition of these maps with the caption plot and the LDA provides a highly specific graphic tool to analyze the intersection of selected chemistries with a topic in glass science or a specific property/characterization technique.

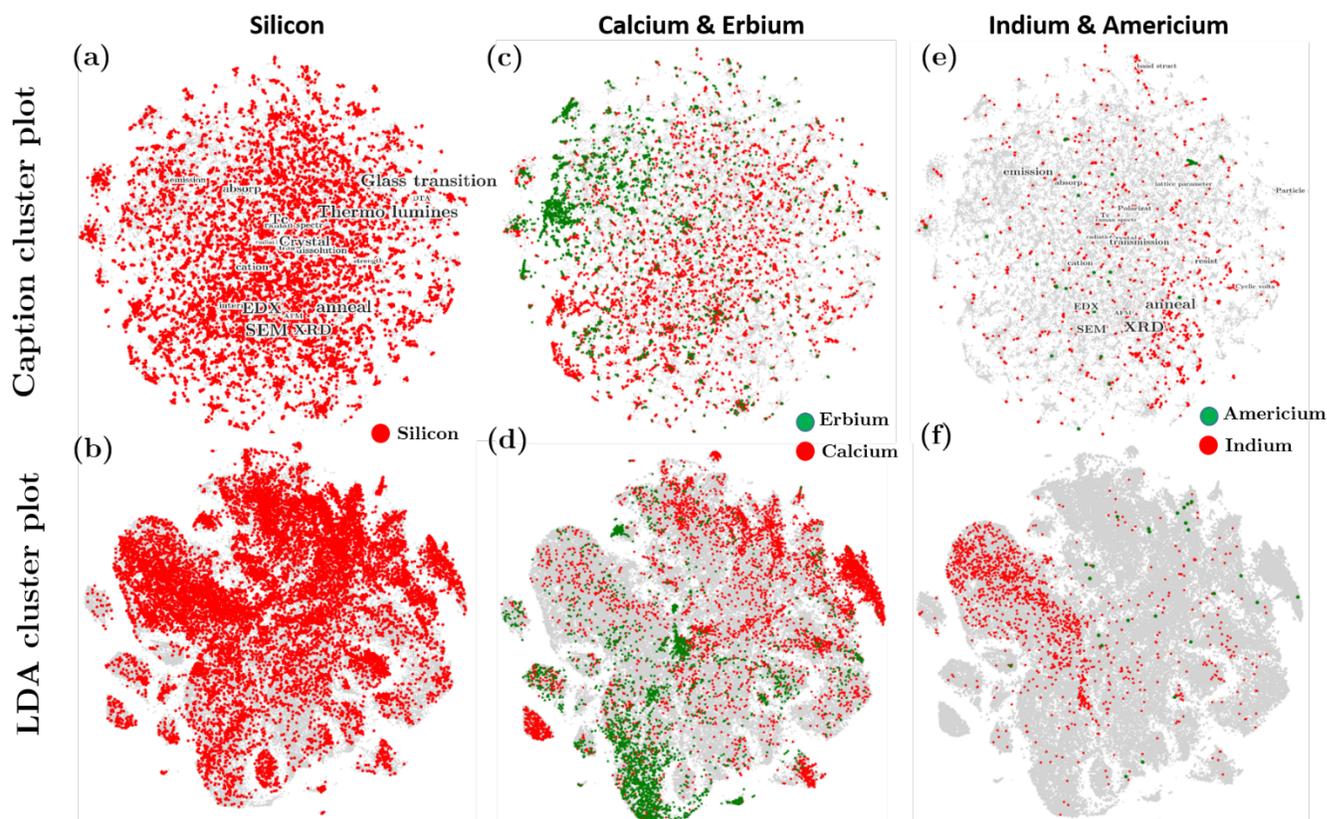

**Figure 3. Elemental maps.** The elemental maps of **(a)** Si, **(c)** Ca and Er, and **(e)** Am and In superimposed on the CCP showing the presence of the respective elements in the abstracts of the manuscripts from which the images are taken. The elemental maps of **(b)** Si, **(d)** Ca and Er, and **(f)** Am and In superimposed on the LDA plot showing the abstracts where the elements have been identified to be present.

As a validation of this concept, it is seen easily from Fig. 3(a) and (b) that silicon is abundantly distributed among all topics, properties, and characterization methods in glass literature. A similar result can be seen for oxygen (see Supplementary material). This is hardly a surprise as silicate glasses are one of the most common family of inorganic glasses. The map for calcium and erbium in Fig. 3(c) and (d) is more illustrative. Calcium is found to be less uniformly distributed than silicon, with high concentration in the clusters identified as bioactive by LDA. This is a representation of the fact that calcium is one of the major constituents of bioactive glasses. The overlapping data for erbium in Fig. 3(c) and (d) shows that this element is mostly present in the LDA topic 'rare earth glasses' and in the caption clusters 'emission', 'PLE', 'band structure' and 'energy diagrams'. The Elemental map therefore provides a visual diagram of the presence of an element in glass literature. Rare earth elements such as erbium, dysprosium and ytterbium are used largely for LEDs and laser applications which is confirmed by the region of the LDA plot that is highlighted by these elements ('Rare earth glasses' and 'LEDs and Phosphors') as well as by the image categories where they predominate.

One of the questions raised in the introduction 'Where is Americium used in glasses' is answered in Figs. 3 (e) and (f). Only 25 articles were found with Americium mentioned in the abstract. They are seen to fall over regions identified as 'glass ceramic' and 'glass irradiation studies' from the LDA plot. Upon inspection, it is indeed found that most of these abstracts relate to studies of nuclear exposure and radiation on glass ceramics – demonstrating a practical use of the concepts developed so far in answering a question that is otherwise not reachable through any other approach today. Similarly, Indium is found to be distributed in Figs. 3(e) and (f) mostly in the section identified as 'Thin films' relating to the large body of work that has been carried out on the transparent conducting Indium Tin Oxide (ITO) electrodes on glass substrates and glass ceramics. The elemental maps for all the 120 known elements overlapping with the LDA and CCPs are presented in the Supplementary material.

**Discussion**
The LDA and CCPs provide a highly specific, detailed, and succinct graphical summary of the available corpus of glass literature. They provide answers to some of the questions raised in the introduction. For example, the caption cluster plot visually conveys the fact that the most studied aspects of glasses in literature relate to their annealing behavior and microstructure, and that studies on optical emission are slightly more common than that of absorption. The database generated through the caption cluster plot provides a useful source of specialized images—say SEM microstructural images or AFM images of glasses—which can then be used for learning images through AI and ML algorithms such as convolutional neural networks. These can be used to answer some of the pressing problems in the field today such as identifying the causes of fracture or dissolution by linking to other structure—processing parameters.

Similarly, the LDA plot provides insights on the broad themes within glass ontology into which the text is divided. It is seen immediately through visual inspection that the amount of work done on bioactive glasses is less than that on glass ceramics or that irradiation studies on glasses occupy the bottom of glass hierarchy in terms of the sheer number of publications. The LDA plot is a topological map of glass literature, where each text is assigned a unique position next to other works of similar nature, content, and theme. This provides a way to efficiently search for publications that are similar to a given paper—once the position of the publication in the vector space is determined, the nearest neighbors are by default the ones that are the most semantically similar. There are currently no other tools, even in established scientific databases and search engines, that allow the detailed exploration, analysis, and easy visual analysis offered by the CCPs and the LDA plots. At the same time, combining the caption cluster, LDA, and Elemental plots results in the creation of a tool that can query, explore, and analyze the information content in glass literature with unprecedented detail and specificity.

An example toward such an attempt for knowledge extraction and dissemination combining CCP, LDA and Elemental map is presented in Fig. 4. The LDA plot identifies abstracts in the database that belong to the category of 'bioactive glasses'. The elemental maps allow the selection of only those abstracts among these that have been marked with the presence of fluorine and chlorine. This is shown in Fig. 4(b) where the red pixels corresponding to the F and Cl containing abstracts are found to overlap with the green pixels corresponding to abstracts on bioactive glasses. The region of the greatest overlap is seen to be the three islands on the LDA plot that are marked as bioactive glasses, thereby confirming that among all fields of glass science, F and Cl most commonly find their application within this topic. This in itself is a remarkable capability—the identification of only those scientific publications subject to the dual constraints of topical category and chemistry. It bears repeating that there is no other method currently available that can do this.

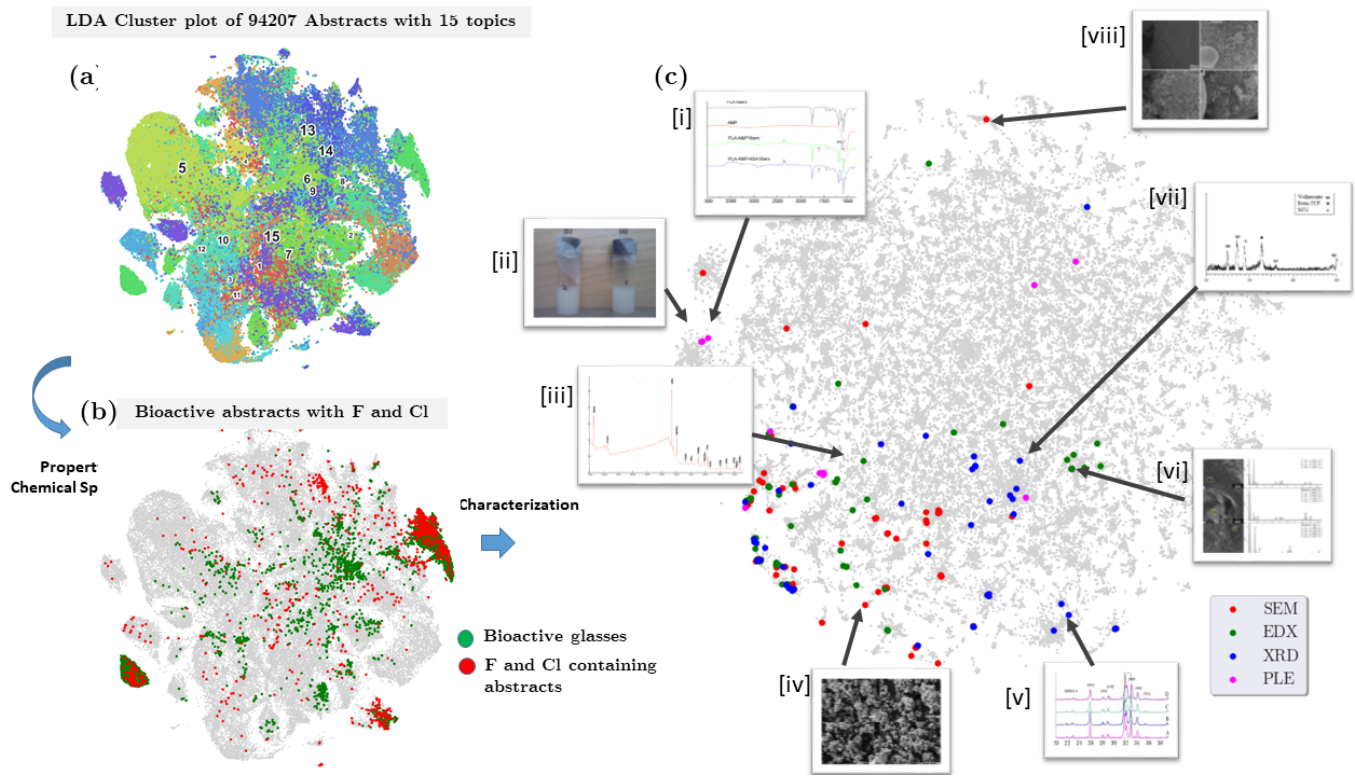

**Figure 4. Knowledge extraction combining CCP, LDA and Elemental maps. (a)** The bioactive glass cluster is identified from the LDA plot. **(b)** The abstracts that contain F and Cl are marked on the plot in red. The area of overlap are abstracts on bioactive glasses that contain F and Cl. **(c)** The SEM, EDX, XRD, and PLE images from journal articles belonging only to this parameter space are marked by colored pixels in the CCP. Inset images [i-viii] are arbitrarily selected examples of these images as identified by the plot[27,35,48-52].

Adding the caption cluster plot to this information allows even deeper exploration of literature by extracting a processing, characterization, or property image from this parameter space. For example, Fig. 4(c) show the SEM, EDX, XRD, and PLE captions of figures from abstracts that contain F and Cl on the topic of bioactive glasses. Arbitrarily selected images from this parameter space are displayed for reference in the inset of Fig. 4(c) [i-viii]. The captions of these images confirm that the figures do correspond to the selected image type, while their abstracts span a broad topic range include apetites, glass microparticles, mesospheres, and bioactive scaffolds—all of which relate broadly to the subset of bioactive glasses under consideration. Thus, this method allows the rapid exploration of scientific data to access extremely nuanced and specific information sets. Such a method might be very useful for a researcher who wishes to access the microstructure of chloride or fluoride glasses without conducting an extensive literature survey—the only alternative available today. Elemental tagging adds a chemical marker to images for computer vision tasks, as in the training of predictive algorithms that link microstructure or property to composition.

The methods of systematic scientific exploration of literature that has been developed so far can be generalized by querying the abstract database for arbitrary search terms. Figure 5 presents an example where the application strings 'optical glass' and 'LED', as well as the synthesis string 'solid state' has been used to identify all the abstracts in the document space that contain the respective strings. This allows us to categorize the abstracts based on specific applications, synthesis methods, characterizations, or properties. In Fig. 5(a), all the abstracts relating to optical glasses are identified from corpus and are visually represented as green pixels in the LDA plot, where they are seen to overlap with the topics identified as 'LEDs and Phosphors', 'spectroscopic studies' and 'rare earth glasses'. Similarly, Fig. 5(b) shows all the abstracts that contain the search string 'solid state synthesis' which is a common method for fabricating oxides and oxide glasses. The overlap of the two sets of complementary information are the set of abstracts on optical glasses synthesized through solid state synthesis. This analysis can be carried further by combining this data with the caption cluster plot, through which specific image types such as the DTA or XRD of optical glasses made through solid state synthesis can be extracted. While the level of specificity offered by this approach is in itself useful to a researcher who wishes to learn more about the solid state synthesis of optical glasses, the method allows the use of any number of search strings—allowing for a detailed multidimensional extraction of information from literature.

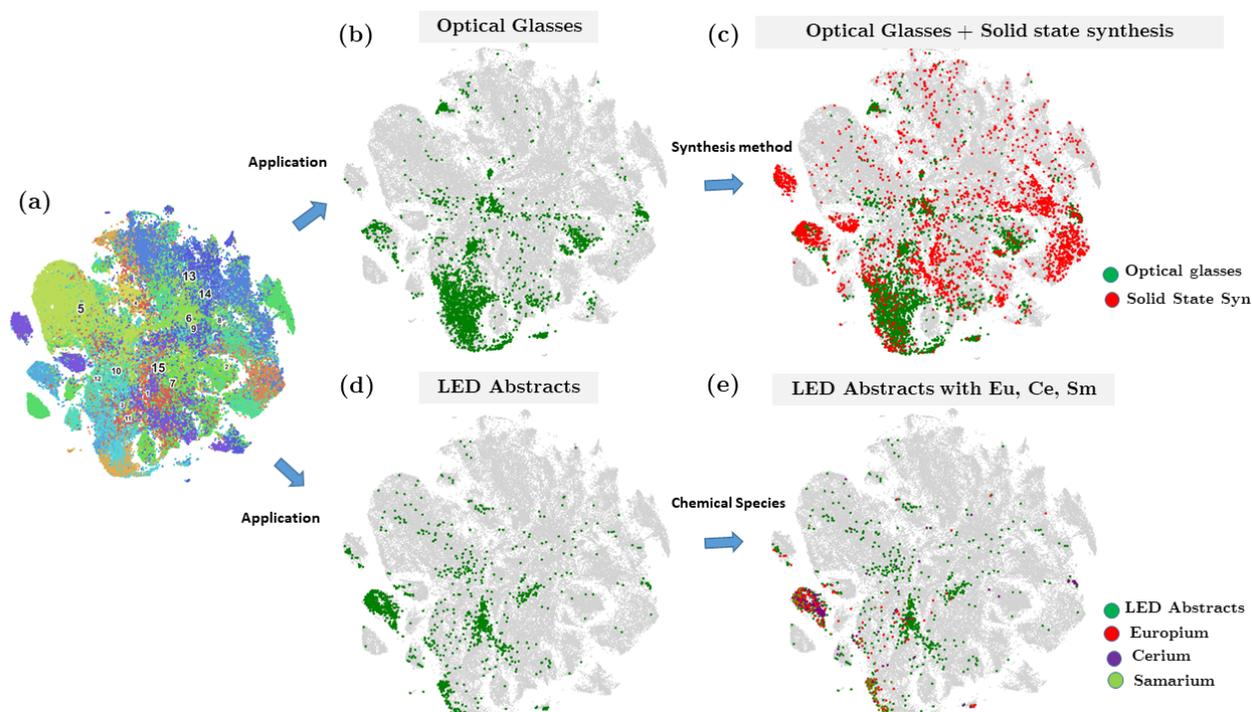

**Figure 5. Knowledge extraction framework. (a)** The LDA plot. **(b)** Abstracts with 'optical glasses' in the text. **(c)** Abstracts with 'solid state synthesis' in the text overlapping with optical glasses. **(d)** Abstracts with 'LED' in the text **(e)** Abstracts with rare earths.

Similarly, the green points in Fig. 5(c) represent abstracts with the string 'LED' in it. By mapping this data to the elemental maps, the presence of abstracts with specific chemical compositions can be marked, such as in Fig. 5(d) that highlights the elements europium, cerium, and samarium. These are therefore LEDs that contain any of these elements or any combination of them. This data allows the user to read through these abstracts, and only these abstracts, to learn more about the subject. He or she is then able to look through microstructural, luminescence, or thermodynamic data that is linked to this dataset through the caption cluster plot.

## Conclusion

Altogether, we demonstrate that the application of NLP to glass literature enables the curation and selection of data sources sorted by specific applications, properties, characterization methods, and chemistries. In turn, this allows for custom composition–processing–property databases to be compiled automatically, and in linking together parts of the information space in a way that has so far not been possible. When used together, text vectorization, LDA, and elemental maps solve many of the challenging questions impeding the accelerated discovery of glasses, some which are stated in the Introduction. However, the major contribution of the present study is that it opens a way to ask deeper and broader questions of glass literature, that can ultimately enrich the field by solving outstanding problems – old and new. In particular, the development of a glass specific ontology, a curated image repository, and a named entity recognition tool tailored for this community can go a long way in enhancing the methods demonstrated for the first time in this paper. Further, by combining these tools with a natural language model—such as a 'GlassBert'—can open the field to many more advances in Artificial Intelligence and Machine Learning by making available rapid, efficient, and tailored composition–processing–property databases. The scalable methods demonstrated in this study are broadly applicable to any scientific field and consequently is of universal relevance in accelerating the scientific enquiry.

## Acknowledgements
N.M.A.K. acknowledges the financial support for this research provided by the Department of Science and Technology, India under the INSPIRE faculty scheme (DST/INSPIRE/04/2016/002774) and DST SERB Early Career Award (ECR/2018/002228). The authors thank the IIT Delhi HPC facility for providing the computational and storage resources.